# Triple-Identity Authentication: The Future of Secure Access

Suyun Borjigin
Independent Researcher
yunsu000@hotmail.com

**Abstract—** In password-based authentication systems, the username fields are essentially unprotected, while the password fields are susceptible to attacks. In this article, we shift our research focus from traditional authentication paradigm to the establishment of gatekeeping mechanisms for the systems. To this end, we introduce a Triple-Identity Authentication scheme. First, we combine each user credential (i.e., login name, login password, and authentication password) with the International Mobile Equipment Identity (IMEI) and International Mobile Subscriber Identity (IMSI) of a user's smartphone to create a combined identity represented as "credential+IMEI+IMSI", defined as a system attribute of the user. Then, we grant the password-based local systems autonomy to use the internal elements of our matrix-like hash algorithm. Following a credential input, the algorithm hashes it, and then the local system, rather than the algorithm, creates an identifier using a set of elements randomly selected from the algorithm, which is used to verify the user's combined identity. This decentralized authentication based on the identity-identifier handshake approach is implemented at the system's interaction points, such as login name field, login password field, and server's authentication point. Ultimately, this approach establishes effective security gates, empowering the password-based local systems to autonomously safeguard user identification and authentication processes.

## 1. INTRODUCTION

In the digital age, multi-factor authentication (MFA) [1], [2] is nearly ubiquitous across online world. As a third-party service, MFA can help confirm a user's identity by providing an extra code. However, MFA serves as an external auxiliary system in which the code is transmitted over the network and then entered manually by users. On the other hand, traditional password-based authentication systems lack robust internal mechanisms to protect their interactions with the outside world. Specifically, username fields typically offer almost no protection, while password fields remain vulnerable due to users' weaknesses in password management. Particularly, the single-factor identity presented by users is verified using hash values generated by cross-system hash algorithms. However, this handshake approach has consistently encountered challenges, such as reverse engineering and collision attacks, resulting in password-based local systems increasingly relying on external assistance.

Typically, access control mechanisms depend primarily on two aspects: user identification and authentication in the background of the local system. Identity representation is pivotal in user identification; however, it has always been the weakest link in the authentication process. Hash values play a crucial role in authenticating users, but their generation relies entirely on cross-system hash algorithms. On the other hand, password-based local systems have seen little progress over decades, with respect to their inherent vulnerabilities. Compared with researches on multi-factor authentication, in-depth studies on the security of the local systems themselves have not received the attention they deserve in mainstream research areas.

This article shifts the research focus to the establishment of gatekeeping mechanisms for the interactions between password-based systems and the outside world. First, we redesign the representation of the user's identities to enable the local systems to accurately authenticate users. Secondly, we develop a novel approach of generating identifiers for the password-based local systems to securely verify the redesigned user identities [1], [3], [4].

First, let us reexamine the security landscape of a successful MFA-based authentication process. In this process, entering the password by a user triggers the MFA system to transmit an additional code to their smartphone over the network. When the received code is entered, the user can then be granted access to their account. In this process, the password served as the first factor to initiate the online transmission of the code. The user's smartphone identified by the International Mobile Equipment Identity (IMEI) is used to receive the code as the second factor. The user's subscription linked to the International Mobile Subscriber Identity (IMSI) makes it possible to transmit the code to the smartphone.

This scenario illustrates that the device identities (such as the IMEI and IMSI numbers), along with the user credentials, are indispensable factors for the security of user identification. However, this success depends on the collaboration of two systems. One is the password-based local system, which is responsible for the initial identification of the user credentials. The other is the MFA-based system, which is responsible for receiving the second factor transmitted through the subscribed network service to verify the user's device.

In views of this, we propose a Triple-Identity Authentication (TIA) that integrates the aforementioned three factors to perform accurate user identification at all interaction points between the local system and external entities, thereby authenticating legitimate user.

In terms of the current user identification paradigm, even though users may create relatively complex passwords with the help of certain technologies, such as password managers, they ultimately present the system with single-factor password that are not genuinely secure. In this article, we design a technology for the local system to integrate the IMEI and IMSI numbers [5-7] of the user's mobile device with their credentials. This approach randomly integrates user's identities, their device, and service into a cohesive structure, forming a multi-factor identity that represents the user's identity. This man-device-service structure is referred to as a combined identity, represented as "credential+IMEI+IMSI", which is created by the local system rather than the user and identified in the background of the system.

Through this integration, the login credentials that represent the user's personal attributes are transformed into the combined identity that stands for the user's system attribute. The former is created by the user in a less secure environment, while the latter is managed by the system in the background. It is evident that this approach shifts the burden of login security from the user to the local system. This makes it possible to reduce the complexity of the credentials, such as containing only numbers and lowercase letters as valid characters. Nevertheless, this combined identity is difficult to replicate on unauthorized devices with different IMEI and IMSI numbers.

In the realm of user authentication, a specific hash algorithm consistently converts the same user input across different local systems into identical hash values [8], which are then utilized to verify user's identities. This deterministic property indicates that the mapping from input to output is fixed, resulting in predictable outcomes. While this deterministic hashing function is intentionally non-random and lacks the properties of randomness, it is important to recognize that these characteristics can potentially be exploited by malicious actors for reverse engineering and collision attacks. Therefore, regardless of the strength of the deterministic hash algorithm, it remains challenging for local systems to safeguard their access points (such as username and password fields) without seeking additional support or measures.

In this study, password-based local systems will substantially participate in the process of authenticating users, thereby overcoming the non-random and deterministic characteristics to ensure the authentication of legitimate user. Specifically, we grant local systems a certain degree of autonomy to utilize the internal elements of the algorithm to randomly generate a unique identifier, and then use it to verify the aforementioned combined identity.

To achieve this, a single-character conversion technique [9-11] is utilized to establish a matrix-like hash algorithm for password-based authentication systems, as shown in Figure 1. This algorithm converts each login credential entered by a user into a matrix of hash elements containing characters in any language that the computer can process. Subsequently, we open the internal structure of the algorithm (i.e., the matrix elements) to all user credentials, such as login name (i.e., username and phone number) and the pair of login and authentication passwords [11], [12], as shown in Figures 1 and 2. Therefore, the local system can randomly select a set of hash elements from the matrix to generate a unique identifier associated with the entered credential, and then utilize it to verify the combined identity. In this study, everything entered into the system through login fields will be hashed by the algorithm, whether secret or not.

As the open algorithm runs in the background, its internal hash elements are concealed within the system, inaccessible to users, independent of user information, and not transmissible in cyberspace. The identifiers made with such elements are ideal for the local system to verify the combined identity. In contrast, they are useless to hackers as they contain invalid characters that are not allowed to enter the system through user interfaces (such as login name field and login password field). In addition, the length of authentication passwords and identifiers is variable, making it extremely difficult to reverse engineer the original credentials.

During registration, users only need to provide their login credentials: a login name and login password through their smartphone. In the background, the system combines each credential with the IMEI and IMSI of the smartphone in specific manners to generate corresponding combined identities. Upon establishing the matrix of the hash algorithm, the identifier and authentication password related to the entered credential are generated. The system stores the credentials, identifiers, and authentication password.

In a login process, entering a username (UN) in the login name field via a user's smartphone initiates an identification process to check the compatibility of the username with the IMEI and IMSI. If they match, a combined UN identity will be generated, followed by the creation of a matrix of hash elements. Subsequently, the system creates a UN identifier using a set of hash elements randomly selected from the algorithm matrix in order to verify the combined UN identity. Upon the handshake verification, the user can proceed to the password entry page. In case the entered login name is a phone number (PN), it also goes through the same process to generate a combined PN identity and a PN identifier to perform the handshake process.

When it comes to the password page, the system checks whether the entered login password (LP) matches the IMEI and IMSI. If compatible, a combined LP identity is generated and then the login password is converted into a matrix of hash elements. Using a randomly selected set of hash elements from the matrix, the system creates a corresponding LP identifier, which is then used to perform the handshake verification. Only after the successful verification can an authentication password (AP) be generated by the algorithm, as illustrated in Figure 1. Similarly, upon creating a combined AP identity, an AP identifier is generated using another set of hash elements randomly selected form the same matrix for the handshake verification on the server. Moreover, the authentication password itself can also serve as an AP identifier to verify the structure of "IMEI+IMSI", representing a streamlined combined AP identity.

By performing the identity-identifier handshake verification on three critical interaction points, the TIA scheme establishes an effective gatekeeping mechanism for password-based authentication systems. As a result, traditional password-based systems can autonomously protect the process of user identification and authentication.

## 2. MATRIX-LIKE HASH ALGORITHM

In this study, we employ a single-character conversion technique [9-11], which randomly converts a single character into a string of characters. Therefore, each character in a password chosen by a user can be converted, resulting in a set of strings.

For example, a user-selected character "d" is randomly converted into a six-character string "3Mo&(E" after the digit "6" is selected from the drop-down menu. This process is illustrated in the conversion unit of the second row, as shown in Figure 1. Subsequently, other characters "p", "7", "a", "3", and "k" selected by the user are individually converted into a set of strings "vX#", "z%9CP", "?G", "d$L", and "Q". These processes occur after the corresponding digits "3", "5", "2", "3", and "1" are chosen from the drop-down menus, resulting in an additional five strings by the relevant conversion units.

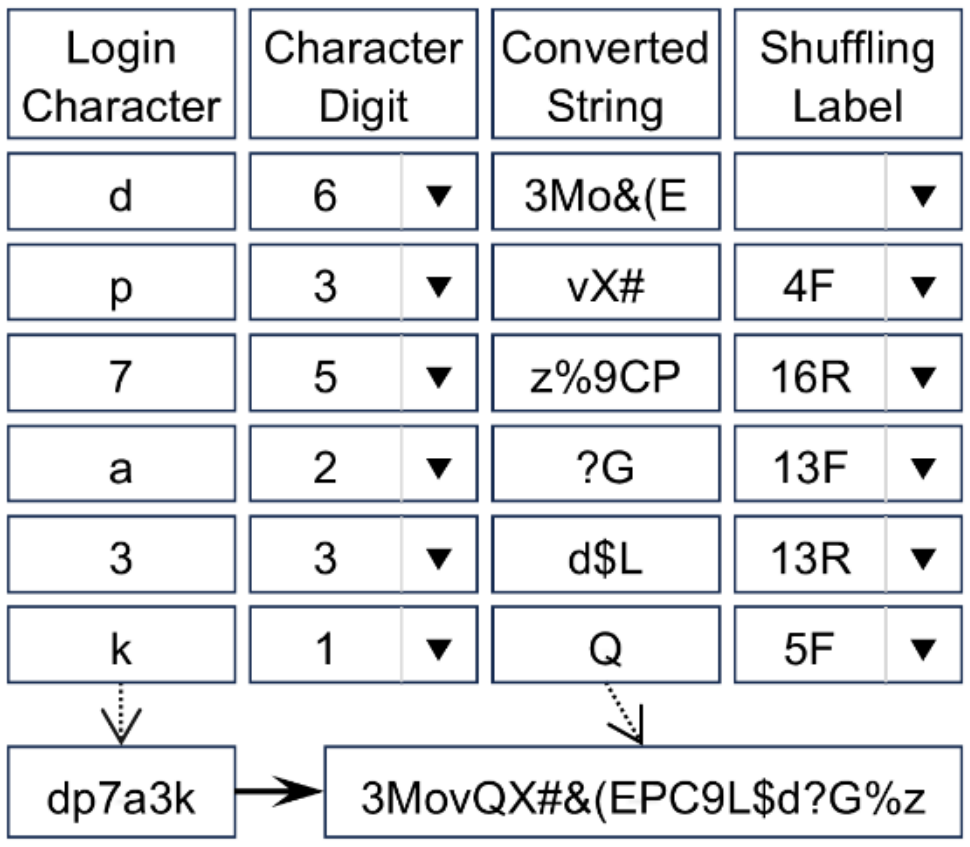

**Fig. 1.** The matrix-like hash algorithm and the conversion of a login password into an authentication password.

### 2.1. Generation of a Matrix-Like Hash Algorithm

By stacking the above six units, a two-dimensional framework is created. Next, a column of instructions referred to as Shuffling Label is attached to the right of the framework, forming a matrix-like structure consisting of six rows and four columns, as shown in Figure 1. The first column is designated as Login Character. The second column is designated as Character Digit, while the third is called Converted String. In this structure, each label indicates that its left-hand string is inserted into the preceding string, in a manner similar to shuffling a deck of cards.

For example, the label "4F" means that the second string "vX#" as a whole is inserted into the fourth insertion point of the first string "3Mo&(E", in a forward character order indicated by the letter "F". This results in a temporary string "3MovX#&(E". The label "16R" is to insert the third string "z%9CP" into the sixteenth insertion point of the temporary string in a reverse character order (i.e., "PC9%z") represented by the letter "R". However, there are only 10 insertion points in the temporary string, which means that the third string can only be inserted into the 10th insertion point, thus generating the second temporary string, represented as "3MovX#&(EPC9%z". When all label instructions are executed in sequence, the original string "dp7a3k" is processed through the matrix-like structure, resulting in a longer and more complex string "3MovQX#&(EPC9L$d?G%z", as shown at the bottom of Figure 1.

It is clear that this matrix-like structure functions as a hash algorithm that converts a string of login characters entered by a user into a longer and more complex string. This structure is referred to as a matrix-like hash algorithm. Subsequently, we integrate this structure into a password-based authentication system, making it the hash algorithm of the system. Once this system in place, users can enter their login characters, such as "dp7a3k", into the password field. This initiates the system to generate a matrix of six rows and four columns of hash elements, and then produce the aforementioned longer string. In this way, all operations, such as the selection of drop-down menus, conversion of strings, and implementation of labels, can be automatically executed by the system in the background without users' participation.

In the context of this system, the string of login characters used to log in is defined as a login password, representing user's personal attribute. The longer and more complex string is defined as an authentication password, which is stored for later use in verifying a user's identity. Together, they build a pair of login and authentication passwords, as shown at the bottom of Figure 1.

### 2.2. Unique Features of the Matrix-Like Hash Algorithm

In addition to the function of hashing a user credential into a complex one, this seemingly simple matrix-like hash algorithm has the following unusual features that can be exploited to their full potential.

1) As a functionality of authentication passwords, it is designed to serve as a hash value, which can be utilized by the local system to verify user's identities. In Figure 1, the sum of a selected set of digits from the Character Digit column is variable, resulting in a variable-length authentication password. The hash values generated by traditional algorithms are fixed in length and unchanging; this determinism may be exploited by hackers for reverse engineering and collision attacks. In contrast, the variable-length authentication password can prevent such malicious activities. The other functions of the authentication passwords will be introduced in Section 6.

2) Generating hash values is the core functionality of traditional hash algorithms. In the authentication processes, user inputs are associated with hash values through cross-system hashing algorithms. However, this linkage can also be utilized by hackers to conduct reverse engineering or collision attacks. In this study, we aim to develop a method of generating hash values by local systems, rather than relying on conventional hash algorithm, to implement user authentication. As illustrated in Figure 1, the local system randomly selects a set of internal elements from the matrix-like hash algorithm and combines them into a unique string, referred to as an identifier. This locally generated identifier has no algorithmic relationship with the user input, so the local system can utilize such identifiers to verify user's identities. Consequently, this approach establishes a novel authentication model. It not only effectively addresses the inherent defects related to the non-randomness and determinism of cross-system algorithms, but also renders reverse engineering and collision attacks ineffective.

## 3. Identity and Identifier in the Authentication System

In the realm of digital authentication, the concepts of "identity" and "identifier" often overlap [13], [14] due to their contextual usage, making them frequently interchangeable in various contexts. In an authentication process, the user's input is hashed by an algorithm into a hash value. The former signifies the user's personal attributes, while the latter represents the user's system attributes managed by the system. Although both refer to the same subject at a fundamental level, they are perceived from different perspectives belonging to distinct subjects, rendering them conceptually disparate. However, in both theory and practice, the two concepts are often used interchangeably, and this conflation of concepts can impact progress in related fields. Thus, it is necessary to clearly define and elucidate the two concepts to enhance our understanding of their respective roles and functions.

User identification and authentication are fundamental aspects of access control. In this study, we think that any data entered into the system through login fields should be classified as "identity". The accuracy and security of user identification depend on the combination of multiple identity factors, as single-factor identification has been proven to be insecure. Instead, the user authentication relies heavily on the randomization of the identifiers. However, existing user identification models still rely on a single factor based on user's personal identity, and thus have to use more factors provided by external MFA services to ensure the identification security. Conversely, user authentication relies entirely on hash values generated by cross-system hash algorithms, which have consistently faced challenges due to algorithm's non-random and deterministic nature. To address these issues, it is crucial to develop a method to present user's identity that includes all integral identity factors. And it is more important to develop a method of generating truly random identifiers that can be used to authenticate users in an environment free from external interference.

Given that the concepts of "identity" [1], [3] and "identifier" [1], [4] encompass multiple interpretations and are often used interchangeably, it is important to provide precise definitions for them. This clarity is crucial for paving the way for generating tamper-proof identities and truly random identifiers. In the following sections, we will explore these concepts in detail, particularly their roles in user identification and authentication.

### 3.1. Identification Factors: The User's Identity

**1) Identity in the Context of This Study**: Traditionally, a username is classified as the identity of a user, serving as a unique label that distinguishes one individual from another. On the other hand, the password is viewed as an identifier to validate that identity and grant access to the associated resources. In this study, we focus not on the classification of the data entering the system, but rather on the overall security framework that protects it. Since the system interacts with the outside world primarily through login fields, it is essential to establish a robust security mechanism at each interaction point to ensure that all system interfaces can withstand potential threats.

Typically, user-created identities are characterized by their simplicity, often relying on a single factor for authentication, which represents a significant flaw. This suggests that the solution lies in combining as many integral user identities as possible to present a more comprehensive representation of the user. However, achieving this integration requires the application of certain technologies that facilitate identity management and verification.

In essence, identity serves as evidence of an individual's self-identified persona, thereby facilitating the process of user identification. In addition, identity can also function as a mechanism for establishing proof of ownership for a specific object, thereby facilitating the identification process of that object [15]. If the object belongs to the user, then its identity is regarded as a part of the user's identity family. Therefore, user's identities can be conceptualized as a large family that includes a diverse array of physical and digital identities within the real world. Example include license plates, email addresses, passport numbers, digital certificates, user IDs, user credentials, and IMEI and IMSI numbers [6], [7], [13] among others.

In this study, we particularly focus on the identities associated with users, their mobile devices, and the online services to which they subscribe. Users are commonly identified by their login names, such as usernames or phone numbers, along with their passwords. This indicates that the traditional role of passwords has evolved from serving as identifiers to becoming vital components of identity. Additionally, users' mobile devices (such as smartphones) and the services they subscribe to are associated with the IMEI and IMSI numbers, respectively. However, it is important to note that users typically do not interact directly with these numbers; instead, they operate seamlessly in the background, managed by the local system.

**2) Multi-factor identification**: In a typical login process, the local system is unable to ascertain whether a login attempt based on a single factor identity is from a device recognized as legitimate and registered to the user. To enhance security, an external MFA service is employed to send a unique code to the user's mobile device, serving as the second factor of identification. Upon entering the code, the local system can recognize the device that received the second factor as "trusted", thereby allowing access to the user's account. In this successful login process, the login credentials, IMEI (related to the user's device), and IMSI (pertaining to the mobile service) are indispensable components, playing a crucial role in strengthening user identification.

Based on this scenario, we integrate these identities of the user: the credentials, IMEI, and IMSI into a single framework to represent the user's identity. This framework is defined as a combined identity and denoted as "credential+IMEI+IMSI". In this way, all essential components required for a successful login are encompassed within this trinity identity framework.

In practical applications of the combined identity, inputting a credential in the designated field initiates a verification process within the local system to assess the compatibility of the credential with the IMEI and IMSI of the user's smartphone. Once verified, the system generates a combined identity associated with the entered credential. Following this, the system authenticates this identity against the corresponding credential identifiers (to be discussed in the next section).

For example, when a user enters their username (UN) in the login name field, the system begins by verifying its compatibility with the IMEI and IMSI. Upon successful verification, the system generates a combined UN identity for further authentication. Once this identity is authenticated, the user can enter their login password (LP) in the designated field. After confirming the password's compatibility, a combined LP identity is generated for authentication purposes. Only after this step can the algorithm create an authentication password (AP). Finally, the server generates a combined AP identity for authentication following a successful verification of its compatibility.

Typically, login credentials are considered to represent the personal attributes of the user, which are single-factor identities. Conversely, the combined identity herein stands for the system attribute of the user, which is a multi-factor identity. Considering the difficulties users face in creating robust passwords and the shortcomings of single-factor identities, this study transforms the single-factor personal identities into multi-factor system identities, which are generated by integrating users' credentials, devices, and services. Although the latter still belongs to the user, it is managed by the system in the background, rather than by users in a less secure environment.

This integration provides a multi-factor identification mechanism for local systems, which includes all the essential identity elements necessary for a successful login. The incorporation of the IMEI into the identity framework serves a critical function by enabling local system to accurately identify the specific mobile devices from which users input their credentials. In addition, the involvement of the IMSI allows the systems to verify that the user entering the credentials in indeed a legitimate registered user, rather than an unauthorized individual. Therefore, this multi-factor identity makes it impossible for unregistered users to successfully present the system identities of legitimate users on devices with different IMEIs, thereby accessing restricted resources.

**3) Complexity of the combined identity**: Typically, character length is crucial to the strength of a user's identity. Authentication systems often require the user identities to be as complex as possible for satisfying the security requirements. However, for the vast majority of users, this presents a significant challenge, leading to a notable decline in usability for clients. This highlights the inherent trade-off issues faced in balancing usability and security. In the TIA framework, the security of user identification and authentication mainly lies in the strength and randomness of the combined identity "credential+IMEI+IMSI"

First of all, the successful login based on MFA reveals a fundamental principle: integrating as many identity as possible is crucial for the security of user identification. Within the composition of the combined identity, the IMEI and IMSI are 15 digits long, totaling 30 digits. When they are combined with a user's credential, the length of the combination should be greater than 35 characters. This strength is sufficient to meet password strength requirements.

More importantly, it is crucial that user identity is authenticated in a secure environment within the system. And the three identity factors are randomly combined by the local system rather than the user in a predefined group salting manner, and this combination of more than 35 digits long is managed by the system in the background. In such a securer environment, the security of user identification undoubtedly can be ensured.

A combined identity with such strength and security, which managed by the local system, can definitely meet the system' security requirements. A closer look at this identity reveals that the strength of the credentials is no longer as important as it used to be, which makes it possible to simplify it appropriately to meet the needs and preferences of the average user inherently for usability. In Section 6, we will focus on ways to simplify credentials and ultimately address trade-off issues.

### 3.2. Authentication Factors: The System's Identifier

**1) Decentralized identifiers**: Mainstream authentication methods rely primarily on the hash values generated by cross-system hash algorithms to verify the user identity. While these algorithms play a critical role in authentication, their inherent non-random and deterministic nature presents vulnerabilities, which can be exploited by hackers to reverse-engineer the entered passwords. In this cross-system algorithm-dominated authentication model, the local systems are not authorized to participate in the process of generating hash values (i.e., identifiers), thereby making the authentication consistent across all local systems.

Let us examine the contents of the algorithm matrix depicted in Figure 1. This matrix contains uppercase and lowercase letters, numbers, and symbols, and can also include characters from any language. Consequently, it offers an extensive range of randomness, making these elements ideal for constructing identifiers. Therefore, we empower the local system with autonomy to randomly select a set of elements from the matrix-like algorithm to combine into a string, which we define as an identifier. This identifier is then associated with the combined identity of the corresponding credentials, allowing the system to utilize it for user authentication.

This method constitutes a decentralized mechanism for generating hash values, which fundamentally transforms the traditional approaches to user authentication. By establishing an identity-identifier framework that utilized locally generated identifiers as hash values, this novel mechanism enables the verification of a user's system identity in a unique way. As a result, password-based local authentication systems can be no longer dependent on algorithm-generated hash values for identity verification. Instead, they can autonomously perform user authentication by employing highly random identifiers created locally. This shift not only enhances security but also streamlines the authentication process.

**2) Characteristics of the identifiers**: In this study, an identifier serves primarily as a hash value to enable the local system to verify the combined identity of the user. Unlike traditional hash value generation methods, this identifier is generated by the local system using the internal elements randomly selected form the matrix-like hash algorithm. This decentralized identifier inherently possess a greater potential for randomness compared to those derived from deterministic process of conventional algorithms.

As illustrated in Figure 1, the converted strings are randomly generated following a randomization process of selecting the drop-down menus. When the algorithm matrix, characterized by its high degree of randomness, is integrated into the local system, those internal elements (excluding the Login Character column) are totally concealed from user. Moreover, the highly randomized identifiers themselves do not contain personal information and are not connected to user input through algorithms, like the traditional link between hashes and user inputs. Additionally, these identifiers do not need to be transmitted over the network, so users do not need to manually input them into the system.

Due to the concealment, the identifiers remain inaccessible to individual users, thereby minimizing the risks associated with manual data entry and enhancing the uniqueness of each identifier. Moreover, the absence of personal information in the identifiers significantly reduces the likelihood of identifier theft, thereby protecting individuals' privacy. The feature of non-transmissibility further enhances security, as it eliminates the need for identifiers to be shared over the network, preventing unauthorized replication or sharing. Additionally, the identifiers are capable of containing characters from any language and varying in length, which not only broadens their applicability but also strengthens the aforementioned security features.

**3) Association with the decentralized identifiers**: In the composition of the identity "credential+IMEI+IMSI", only the credentials are variable components. When an identifier is generated using the internal elements from the algorithm matrix in Figure 1, it is bound to the identity, which means that it is also associated with the credential. Unlike traditional association mechanism, the identifier is linked to the more secure combined identity rather than the entered credentials created by users.

This association ensures that even a minor change in the credential will not only yield a drastically different identifier but also lead to significant changes in the combined identity. Therefore, the identifiers can effectively represent and encapsulate user's identities, such as the credentials, IMEI, and IMSI, while concealing the actual content.

In a login process, if the compatibility check of the credentials is successful, a combined identity will be generated. Consequently, the entered credential is converted into an algorithm matrix, which is then used by the local system to create an identifier. When a credential undergoes an unauthorized change, its credential compatibility with the device and service will be compromised. Therefore, the system refuses to generate the algorithm matrix for the credential, thus unable to generate the corresponding identifier.

In this decentralized mechanism, some key functions of traditional cross-system algorithms have been shifted to local system. The highly randomized identifiers break the traditional algorithmic relationship between hash values and user inputs, thereby completely eliminating the chances of reverse engineering and collision attacks where two different users might inadvertently receive the same credential. Moreover, the decentralized identifiers are not only associated with user credentials, but also linked to their devices and services. This method, which integrates all essential factors within a single system, avoids the potential weaknesses of existing authentication methods, such as the transmission of the second factor over cyberspace and manual input of that factor.

## 4. Gatekeeper Mechanism for System Interactions

The matrix-like hash algorithm, as shown in Figure 1, serves two primary functions. The first function is to hash the login password into an authentication password, akin to traditional hashing methods. The second function allows the local system to autonomously utilize the internal structure of the hash algorithm for generating identifiers; in this sense, it is referred to as an open algorithm. This latter function is the primary focus of this study.

To achieve this, we hash every login credential, such as a login name (i.e., username and phone number) and login password, into a matrix of hash elements. Subsequently, the local system randomly selects a set of hash elements from the algorithm to create an identifier for identity verification. Take a login password (LP) as an example. Once entered into the login password field, it is hashed into a matrix of hash elements, and then the system creates an LP identifier using the selected hash elements. This identifier is employed by the local system to verify the combined LP identity introduced in the previous section. In the same way, the identifiers corresponding to the login name can also be generated for respective identity verification.

This approach establishes an authentication mechanism that pairs the combined identity (i.e., the user's system attribute) with an identifier created by local systems rather than algorithms. The local system employs this approach at its interaction points, such as the login name field, the login password field, and the authentication point on the server. Collectively, these identity-identifier approaches form an internal gatekeeping mechanism for password-based authentication systems, which can autonomously protect the process of user authentication linked to each user credential.

Based on the above analysis, it is clear that our research focus has shifted from algorithm-driven authentication models reliant on centralized encryption frameworks to a locally driven gatekeeping mechanism for authentication. This mechanism effectively mitigates the adverse effects of system interactions with external entities on the authentication process. By adopting this approach, we establish a fully independent and robust security framework for password-based local systems, thereby enhancing authentication security.

Identity impersonation and remote attacks present significant threats to authentication processes. The identity impersonation primarily concerns the manner in which users present their identity, while the remote attacks pertain to the internal authentication mechanism on servers. Although direct control over remote attacks is not feasible, various measures can be implemented to enhance the security of our systems. This underscores the crucial need to strengthen the local system's capabilities to effectively safeguard user identification and authentication without relying on external assistance. In this context, the implementation of a gatekeeping mechanism that employs the identity-identifier approach emerges as a promising solution. In the subsequent sections, we will explore the specific methods for generating the combined identity and the corresponding identifier.

## 5. Client-Side Structure: Redesigning the User Identity

The creation and storage of complex passwords present significant challenges for users, leading many individuals to rely on password managers. Although employing a password manager can be an effective strategy for managing user credentials, entrusting the security of all accounts to a single master password introduces potential risks. The sensitive data contained within the managers represents the users' true identities, rendering it a primary target for hackers.

Additionally, the practice of inputting user identities into login fields presents inherent security vulnerabilities.

Moreover, traditional user identification methods typically involve identifying a user's single-factor personal identity as their true identity, while the identities linked to their device and service are provided by external MFA services over the network. Although the provided credentials are encrypted by hashing algorithms, the inherent weaknesses associated with the algorithms continue to pose challenges for the identification and authentication processes.

To address these issues, we have developed a technology that integrates the essential identity factors of users into a unified architecture, which is then presented to the system. This multi-factor identity is verified using the relevant identifier created by the local system rather than hash algorithms, thereby enhancing the security of the user identification and authentication process.

### 5.1. Trinitarian Identity Framework of the Mobile Login

In today's Internet era, the vast majority of users leverage their mobile devices to access various web services, making mobile login the preferred method for user identification. Typically, these mobile logins are implemented through multi-factor authentication (MFA) services to guarantee the security of user identification.

Technically, an MFA-based mobile login consists of three essential factors: a user's login credentials, a mobile device, and a subscribed online service. The login credentials (i.e., login name and login password) represent the user's personal identity managed by user themselves. The subscribed online service for sending a code to the mobile device is recognized through its IMSI number, signifying the user's service identity. Meanwhile, the mobile device for receiving the code is characterized by its IMEI number, indicating the user's device identity. Among these factors, the absence of the credentials suggests that the login process may not have been initiated yet. The lack of IMSI number implies that the transmission of the code is not feasible. Furthermore, the absence of IMEI number indicates that there is no available device to receive the code.

In summary, the collaborative integration of all necessary identity factors is crucial for successful mobile login. In this study, the user's login credentials (such as login name and login password) and their device and service identities (IMEI and IMSI) are integrated into a combined identity according to a specific manner predetermined by the local system. This approach unifies the user, their device and service to establish a multi-dimensional identity framework, which is referred to as a trinitarian identity "credential+IMEI+IMSI".

While the concept of a trinitarian identity may initially appear complex, it is completely managed by the local system in the background. The purposes of this approach are to significantly enhance the security of the user identification process and to streamline the user's interface operations without compromising security. The latter will be discussed in detail in the next section.

### 5.2. Generation of the Combined Identity

In this study, the term "user credentials" refers to a login name and a pair of login and authentication passwords. The login name may include an email username, phone number, and any user-customized text username. In the case of entering the username (UN), a combined UN identity is created in a predetermined manner by the system and expressed as "UN+IMEI+IMSI". Similarly, for the phone number (PN) of the user's smartphone, a combined PN identity is generated and formatted as "PN+IMEI+IMSI". Moreover, the pair of passwords corresponds to a combined LP identity and a combined AP identity, respectively. These identities are stored in the database during registration for future verification.

When a username (UN) is entered in the login name field, the system first identifies its compatibility with the IMEI and IMSI. Upon successful identification, these components are combined to generate a combined UN identity represented as "UN+IMEI+IMSI". Consequently, the user's personal attributes associated with the username are transformed into system attributes. Following the login password input in the designated field, the system checks for compatibility, subsequently generating a combined LP identity in the form of "LP+IMEI+IMSI". Only at this stage can the login password be converted into an authentication password (AP) through the open algorithm. A corresponding combined AP identity is then created as "AP+IMEI+IMSI".

The primary purpose of the aforementioned configurations is not to fundamentally change the essential role of user credentials, but rather to leverage their intrinsic properties and functionalities in order to establish additional layers of protection for user identification within password-based authentication systems.

### 5.3. Benefits of the Combined Identity

The benefits of the combined identity framework are manifold. First and foremost, the integration of login credentials, IMEI and IMSI revolutionizes the traditional method of presenting user's single-factor personal identity to local systems for identification. This is transformed into a model that verifies the user's system-created multi-factor system identity created within the system. In this identification and authentication model, the use of the trinitarian multi-factor identity (i.e., the combined identity) ensures that only legitimate users can be accurately identified. Furthermore, authenticating this identity with a highly randomized identifier guarantees that only these legitimate users are granted access to restricted resources.

In this context, the trinitarian identity model not only effectively thwarts the initial steps of remote attacks by preventing impersonation from unauthorized devices, but it also ensures that even if any one of the identity factors is compromised, hackers remain unable to establish the legitimate user's combined identity via their devices, unless they have physical control over the user's smartphone.

Secondly, this trinitarian combined identity can also effectively defend against SIM swapping attacks [16], [17]. In such malicious scenarios, an attacker convinces a mobile phone carrier to switch the victim's phone number to a new SIM card embedded in the attacker's device. As a result, the victim may lose access to essential accounts tied to their phone number, leading to potential financial loss and privacy concerns. However, the implementation of a trinitarian identity can provide enhanced security measures, as it prevents the successful identification of a legitimate user's combined identity on any device that possesses a different IMEI. This effectively addresses the risks associated with SIM swapping.

In fact, the TIA system adopts a "trust no entry" philosophy to rigorously identify and verify every input into the system. By adhering to the zero-trust principle [18], the trinitarian identity approach treats all access attempts as potential threats, thereby ensuring that security remains a top priority. Within this trinitarian architecture, user privileges are minimized to allow access only at the level of the IMEI-associated device and IMSI-registered service. Consequently, the identities of the user, their device and service are simultaneously verified at each interaction point in the system, significantly enhancing security during user identification.

Thirdly, integrating a login password into the "IMEI+IMSI" structure creates a complex combination, which makes it possible to considerably reduce the complexity of the login password without compromising security. Thus, we can simplify the login password requirements (to be discussed in detail in the next section) while still maintaining the strength (i.e., length and character type) [19], [20] of the combined identity, thereby providing users with a significant level of user experience and password usability.

## 6. Configuration of the Password Pair and Its Consequences

Typically, the client side prefers secure usability, while the server requires usable security. Practical experience has shown that balancing password usability and security on a single-factor identity poses significant challenges. The solution to address this issue primarily relies on the user side. The TIA system sets two identities: the login password, describing the user's personal attributes, and the combined identity, reflecting their system attribute. Resolving the balance issue depends on the login password.

**1) Configuration of password strength**: Considering the complexity of the combined identity, there is no need to impose stringent requirements on the complexity of the user's login password. In this study, the login password is specified in a range from five to fifteen characters in length and contain only lowercase letters and digits, defined as valid characters. In contrast, the authentication password is required to be at least twenty characters long and must contain four-character classes, such as uppercase and lowercase letters, digits, and symbols [19-21]. Furthermore, a login name, which may encompass an email username or phone number, is generally composed of alphanumeric characters or digits. During hashing, all uppercase letters (if any) will be converted into lowercase to comply with the requirements for login passwords.

**2) Solution to trade-off issues**: In the TIA system, the shift towards combined identities in password complexity not only provides users with an ideal secure usability but also represents the optimal user-friendliness that a text-based authentication system can offer. This approach adeptly satisfies users' needs and preferences without compromising the system's overall security. In contrast, the authentication password generated by shuffling the converted strings in Figure 1 can include any characters from any language that a computer can process. This configuration allows the use of the strings to create unique identifiers that cannot be entered into the system through login fields. As a result, these configurations guarantee that the criteria for usable security are met for users, while simultaneously satisfying stringent security requirements on the server side. This resolution effectively addresses the long-standing inherent trade-off conflict between secure usability and usable security [22], [23], allowing for a harmonious coexistence of both password usability and security.

**3) Best user experience**: The utilization of lowercase letters and digits as input characters significantly enhances user accessibility across various keyboard types. In addition, it is crucial to provide convenient screen operations for modern people depending on mobile devices with small screens. This design choice aligns with principles of secure usability, facilitating seamless and convenient login experiences without compromising security. Moreover, the integration of multiple identification factors within a single structure effectively streamlines the user' login process, alleviating the need for extra codes from external services for authentication. These features collectively contribute to an exceptional user experience (UX) [24], showcasing the user-centered design philosophy inherent in the TIA system. By prioritizing user needs and preferences, the system not only ensures security but also fosters a sense of ease and satisfaction in its operation.

**4) Solution to password reuse**: The reuse of passwords [21] across multiple websites presents a persistent challenge in the domain of digital security. This issue arises because the user's single-factor password is directly verified by systems as a representation of their identity during authentication. However, within the TIA system, the user's login password is transformed into a combination (i.e., a combined identity) of the password, IMEI, and IMSI. These identity components are processed by the local system through a technique known as group salting, and generate the combined identity implemented to verify the user's true identity. As discussed in Subsection 3.1, the complexity inherent in this combined identity results in different outputs, even when the same password is reused across multiple accounts. Consequently, within the TIA framework, the practice of password reuse does not entail significant security concerns.

**5) Configuration of user interface**: The implementation of the password strength settings makes it possible to restrict the characters allowed to be entered into the system through login fields. This study classifies lowercase letters and digits as valid characters, while others are deemed invalid and thus are not allowed to enter the system through login fields. Consequently, the login fields can be configured to accept only lowercase letters and digits, thereby effectively preventing authentication passwords and identifiers from entering the system as they contain invalid characters. This method not only further enhances user experience by simplifying the login process but also significantly strengthens overall system security and usability by ensuring the integrity of input data.

## 7. Server-Side Structure: Creating Identifiers

Presently, the prevalent authentication method for verifying a user's identity depends on a secret handshake. Typically, the handshake is achieved by using a stored hash value to verify the hashed password. In this process, the input password is associated with its fixed-length hash value in a non-random way: given the same input, the algorithm always yields the same hash value. This inherent non-random and deterministic nature of traditional hash algorithms can be exploited by attackers to reverse-engineer the input password.

Instead of using stored hashes generated by cross-system algorithms to verify hashed passwords, the TIA system is endowed with the autonomy to generate identifiers using internal hash elements of the open algorithm to verify the user's combined identity. Therefore, the TIA's handshake verification is between the combined identity and the identifier created by the system rather than by cross-system algorithms.

### 7.1. Identifiers for Decentralized Authentication

Typically, a specific hash algorithm converts identical user input across different local systems into the same hash values, which are subsequently utilized for identity verification. This feature indicates that the mapping from input to output is fixed, resulting in predictability in the outcomes. Consequently, this deterministic hash function inherently lacks randomness. As a result, its non-random and deterministic nature may be exploited by hackers to conduct reverse engineering and collision attacks.

In the TIA system, we have designed variable-length identifiers for the local system using highly randomized hash elements extracted from the open algorithm. Subsequently, the system utilizes the identifiers that are functioned as traditional hash values, to verify combined identities. In this approach, the identifiers play a crucial role in the triple-identity authentication process.

Figure 1 illustrates two rounds of randomization applied to the login password. The first round involves selecting digits from the Character Digit column to randomly generate a set of converted strings. In the second round, shuffling labels are randomly selected to create a matrix of hash elements. Moreover, when the algorithm's internal structure is made available to the local system, it can randomly select matrix elements to generate an identifier for user authentication.

This final step establishes a decentralized authentication mechanism, which fundamentally transforms traditional authentication models that rely on algorithm-generated hashes. Now, user identities can be verified using locally generated variable identifiers rather than hash values generated by cross-system algorithms. More importantly, this mechanism disrupts the traditional associations between hashes and user's identities, rendering reverse engineering and collision attacks ineffective.

In summary, the involvement of the local system in the authentication approach not only effectively addresses the shortcomings of traditional authentication models but also significantly enhances the security of user authentication.

### 7.2. Credential Conversion and Identifier Definition

When entering a login name during the login process, users typically provide a public login name, e.g., a username or phone number. This practice underscores a crucial vulnerability, namely that the login name field is not sufficiently protected. To address this issue, we propose a novel approach that employs the open algorithm to convert the login name into a matrix of hash elements. Subsequently, the local system is granted the freedom to utilize the internal elements of the algorithm to generate unique identifiers associated with the input.

Take a virtual email address "Benz428@woxinet.com" as an example. During the registration, once the username (UN) "Benz428" is entered, the system converts it into a matrix of hash elements, as shown in Figure 2. Following this conversion, the local system can randomly select a set of elements from the matrix to generate a string in a group salting manner determined by the system. This string is defined as a UN identifier and then associated with the combined UN identity described in Section 5. In practice, while users may prefer to enter their complete email addresses, the system only hashes the username, converts uppercase letters (if any) into lowercase, and eliminates any non-alphanumeric characters. As a result, an identity-identifier handshake approach is established for the login name field.

| Username | Character Digit | Converted String | Shuffling Label |
|---|---|---|---|
| B | 3 ▼ | y]Q | ▼ |
| e | 5 ▼ | #ws%8 | 5F ▼ |
| n | 3 ▼ | O6& | 9R ▼ |
| z | 2 ▼ | $d | 17R ▼ |
| 4 | 3 ▼ | )Lh | 13F ▼ |
| 2 | 3 ▼ | zF= | 8F ▼ |
| 8 | 1 ▼ | m | 11F ▼ |

**Fig. 2.** The conversion of the username of the virtual email Benz428@woxinet.com and the internal elements that can be randomly selected to generate a username identifier by the local system.

For example, the local system randomly selects a set of hash elements, such as "4" in Username column, "O6&" in Converted String column, "17R" in Shuffling Label column, "2" in Character Digit column, and "zF=" in Converted String column again. Then, the system combines them together into a single string "4O6&17R2zF=", defines it as a UN identifier, and associates it with the corresponding combined UN identity "UN+IMEI+IMSI". Clearly, the length of the UN identifier is not fixed, depending on the random selection of the hash elements. If the input is a phone number (PN), the algorithm maps it into a matrix of hash elements. Following the mapping, the local system randomly selects a set of elements from the algorithm, combines them into a PN identifier, and then associates it with the combined PN identity "PN+IMEI+IMSI".

When it comes to the login password field, upon hashing the entered login password (LP), the local system randomly selects a set of hash elements from the generated matrix, as shown in Figure 1, to create an LP identifier. The system then utilizes it to verify the combined LP identity "LP+IMEI+IMSI", just as it does for the username. Consequently, an identity-identifier handshake is created for the login password field. Furthermore, following the generation of the authentication password (AP) on the server, the system generates an AP identifier using another set of elements randomly selected from the same matrix, and then utilizes it to verify the combined AP identity "AP+IMEI+IMSI". This approach establishes a robust identity-identifier handshake for the server's authentication point.

Alternatively, the created authentication password can also serve as a traditional hash value, which can be used to verify the streamlined AP identity, such as "IMEI+IMSI".

### 7.3. Benefits of the Unique Identifiers

The standout feature of the handshake mechanism based on the locally generated identifiers lies in its provision for the local system to autonomously utilize the algorithm's internal structure to generate an identifier, which in turn is crucial for verifying the combined identity. This innovative approach, which does not rely on hash values generated by cross-system algorithms for identifier creation, significantly enhances the randomness and security of user authentication. As a result, it effectively addresses the issues of predictability and determinism associated with traditional methods. This results in several advantages for user authentication:

First, using variable-length random identifiers means that you may get different identifiers from the same input. This makes the reverse-engineering process more computationally expensive and time-consuming, without imposing any burden on the authentication system.

Secondly, this identity-identifier handshake approach can effectively guarantee the security of user authentication by equipping each interaction point of the system with a robust gatekeeping mechanism. This novel approach enables traditional password-based authentication systems themselves to provide comprehensive, multi-layered protection for user authentication.

Thirdly, the system-managed identifiers do not need to be transmitted through the network, nor do they need to be manually entered into the system through login fields. This effectively eliminates the risk of information transmitted over the network being intercepted and reduces the likelihood of being compromised by malware that exploits manual input. By implementing this method, the overall security of the local system is significantly enhanced, thereby effectively minimizing its attack surface.

Lastly, due to the robust identity-identifier handshake approach, traditional password-based authentication systems will be able to protect the user authentication without external assistance.

## 8. Verification at the System's Interaction Points

In the TIA system, the login process starts with entering a username (UN) into the login name field, where the system checks whether the input matches the IMEI and IMSI of the user's smartphone. Once successfully identified, a combined UN identity is created and represented as "UN+IMEI+IMSI". If the identification fails, the system is unable to create a combined UN identity. Following the successful identification, the user may proceed to the password entry page. Similarly, if case a phone number is entered, the system implements a similar user identification and authentication process.

Upon the input of a login password (LP) in the password entry page, a combined LP identity "LP+IMEI+IMSI" is created after the input compatibility with the IMEI and IMSI. The system then verifies this identity using an LP identifier created with the hash elements selected from the algorithm matrix. Only after completing this round of identity-identifier handshake can the algorithm generate an authentication password (AP). Consequently, the login process continues with user identification and authentication on the server.

The final round of identity-identifier handshake on the server involves generating a combined AP identity, represented as "AP+IMEI+IMSI", after establishing the identity compatibility. Additionally, an AP identifier is created using another set of hash elements selected from the same matrix. In case the generated authentication password serves as a hash value, the system can utilize it to verify the streamlined AP identity, such as "IMEI+IMSI". As a result, users can be seamlessly granted access to their accounts in either case.

## 9. Conclusion

In the triple-identity authentication (TIA) framework, the password-based authentication system transforms single-factor identities created by users into multi-factor identities managed by the system in the background. This approach addresses the vulnerabilities faced by users, including the challenges of managing their identities and the need for manual identity inputs. As a result, users can confidently interact with the local system, free from concerns that their identities might be compromised on unauthorized devices, thereby enhancing the protection of user identification.

Moreover, rather than relying on cross-system hash algorithm, the local system randomly generates unique identifiers to verify users' multi-factor identities. This method effectively tackles issues related to predictability and determinism in cross-system hash algorithms, rendering reverse engineering and collision attacks ineffective. Within this decentralized authentication framework, users' multi-factor identities can be securely verified, ensuring robust protection for user authentication.

Consequently, password-based TIA authentication systems can autonomously protect the security of the user identification and authentication processes.